\documentstyle[aps,preprint,eqsecnum,floats,epsf]{revtex}
\tighten
\draft
\begin{document}

\preprint{MAD-NT-98-04}
\title{Algebraic Nature of Shape-Invariant and Self-Similar Potentials}
\author{A.~B. Balantekin\thanks{Electronic address:
        {\tt baha@nucth.physics.wisc.edu}},
        M.~A. C\^andido Ribeiro\thanks{Electronic address:
        {\tt aribeiro@nucth.physics.wisc.edu}},}
\address{Department of Physics, University of Wisconsin\\
         Madison, Wisconsin 53706 USA}
\author{A.~N.~F. Aleixo\thanks{Electronic address:
        {\tt aleixo@nucth.physics.wisc.edu}}}
\address{Department of Physics, University of Wisconsin\\
         Madison, Wisconsin 53706 USA,\\
         Instituto de F\'{\i}sica, Universidade Federal 
         do Rio de Janeiro, RJ - 
         Brazil\thanks{{\tt Permanent address.}}}
\date{\today}
\maketitle

\begin{abstract}
  Self-similar potentials generalize the concept of shape-invariance
  which was originally introduced to explore exactly-solvable
  potentials in quantum mechanics. In this article it is shown that
  previously introduced algebraic approach to the latter can be
  generalized to the former. The infinite Lie algebras introduced in
  this context are shown to be closely related to the q-algebras. The
  associated coherent states are investigated.

\end{abstract}


\newpage
\section{Introduction}

Supersymmetric quantum mechanics has been shown to be a useful
technique to explore exactly solvable problems in quantum mechanics
\cite{ref1}. Introducing the function
\begin{equation}
W(x) \equiv
-\frac{\hbar}{\sqrt{2m}}\left[ \frac{\Psi^\prime_0(x)}
{\Psi_0(x)}\right] \,,
\label{sup}
\end{equation}
where $\Psi_0(x)$ is the ground-state wave-function of the Hamiltonian
$\hat H$, and the operators
\begin{equation}
\hat A \equiv W(x) + \frac{i}{\sqrt{2m}}\hat p\,,
\label{opa}
\end{equation}
\begin{equation}
\hat A^\dagger \equiv W(x) - \frac{i}{\sqrt{2m}}\hat p\,,
\label{opad}
\end{equation}
we can show that
\begin{equation}
\hat A\mid \Psi_0\rangle = 0
\label{eqgs}
\end{equation}
and
\begin{equation}
\hat H - E_0 = \hat A^\dagger \hat A\,.
\label{eqhe}
\end{equation}

An integrability condition called shape-invariance was introduced by
Gendenshtein \cite{ref2} and was cast into an algebraic form by
Balantekin \cite{ref3}. The shape-invariance condition can be written
as 
\begin{equation}
\hat A(a_1) \hat A^\dagger(a_1) =\hat A^\dagger (a_2) 
\hat A(a_2) + R(a_1) \,,
\label{eqsi}
\end{equation}
where $a_{1,2}$ are a set of parameters. The parameter $a_2$ is a
function of $a_1$ and the remainder $R(a_1)$ is independent of $\hat
x$ and $\hat p$.
Not all exactly solvable potentials are shape-invariant
\cite{ref4}. In the cases studied so far the parameters $a_1$ and
$a_2$ are either related by a translation \cite{ref4,ref5} or a
scaling \cite{ref6}. Introducing the similarity transformation that
replaces $a_1$ with $a_2$ in a given operator
\begin{equation}
\hat T(a_1)\, \hat O(a_1)\, \hat T^\dagger(a_1) = \hat O(a_2)
\label{eqsio}
\end{equation}
and the operators
\begin{equation}
\hat B_+ =  \hat A^\dagger(a_1)\hat T(a_1)
\label{eqba}
\end{equation}
\begin{equation}
\hat B_- =\hat B_+^\dagger =  \hat T^\dagger(a_1)\hat A(a_1)\,,
\label{eqbe}
\end{equation}
the Hamiltonian takes the form
\begin{equation}
\hat H - E_0 =\hat B_+\hat B_-\,,
\label{eqhb}
\end{equation}
The Lie algebra associated by the shape-invariance is defined with the
commutation relations
\begin{equation}
[\hat B_-,\hat B_+] =  \hat T^\dagger(a_1)R(a_1)\hat T(a_1) 
\equiv R(a_0)\,,
\label{eqcb1}
\end{equation}
and
\begin{equation}
[\hat B_+,R(a_0)] =  [R(a_1)-R(a_0)]\hat B_+\,,
\label{eqcb2}
\end{equation}
\begin{equation}
[\hat B_+,\{R(a_1)-R(a_0)\}\hat B_+] =
\{[R(a_2)-R(a_1)]-[R(a_1)-R(a_0)]\}\hat B^2 \,,
\label{eqcb3}
\end{equation}
and the Hermitian conjugates of the relations given in
Eq.~(\ref{eqcb2}) and Eq.~(\ref{eqcb3}). In general there is an
infinite number of such commutation relations, hence the appropriate
Lie algebra is infinite-dimensional. In some special cases where the
parameters are related by translation it is possible to reduce this
infinite-dimensional algebra to a finite dimensional one
\cite{ref3,ref7,ref7a}. In this paper we explore the relationship
between $q$-algebras and the cases where the parameters are related by
scaling.

\section{Coherent States}

Since the operator $\hat B_-$ satisfies the relation 
\begin{equation}
\hat B_-\mid \Psi_0\rangle = 0\,,
\label{eqbgs}
\end{equation}
and the excited states can be written in the form 
\begin{equation}
\mid \Psi_n\rangle \propto \hat B_+^n\mid \Psi_0\rangle\,,
\label{eqppsn}
\end{equation}
the operator $\hat B_-$ does not have a left inverse and the operator 
$\hat B_+$ does not have a right inverse. However a right inverse for
$\hat B_-$ 
\begin{equation}
\hat B_-\hat B_-^{-1} =  1
\label{eqbei}
\end{equation}
and a left inverse for $\hat B_+$
\begin{equation}
\hat B_+^{-1}\hat B_+ =  1
\label{eqbai}
\end{equation}
can be defined. Similarly in the Hilbert space of the eigenstates of
the Hamiltonian, the inverse of $\hat H$ does not exist, but
\begin{equation}
\hat H^{-1}\hat B_+ =  \hat B_-^{-1}
\label{eqhb1}
\end{equation}
does. Also introducing
\begin{equation}
\hat Q^\dagger =  \hat H^{-1/2}\hat B_+
\label{eqqd}
\end{equation}
and its Hermitian conjugate
\begin{equation}
\hat Q = (\hat Q^\dagger)^\dagger = \hat B_-\hat H^{-1/2}
\label{eqq}
\end{equation}
one can show that
\begin{equation}
\hat Q\hat Q^\dagger = \hat 1\,.
\label{eqqq}
\end{equation}
The normalized excited states can then be written as 
\begin{equation}
\mid \Psi_n\rangle = (\hat Q_+)^n\mid \Psi_0\rangle\,,
\label{eqpsn}
\end{equation}
provided that the ground state is normalized, i.e. $\langle \Psi_0\mid
\Psi_0\rangle = 1\,.$

We introduce the coherent state for a shape-invariant potential as 
\begin{eqnarray}
\mid z\rangle &=& \mid 0\rangle\ + z\hat B_-^{-1}\mid 0\rangle + z^2\hat
B_-^{-2}\mid 0\rangle + \dots \nonumber \\
&=& \frac{1}{1-z\hat B_-^{-1}}\mid 0\rangle ,
\label{eqcs}
\end{eqnarray}
where we used the short-hand notation 
$\mid 0\rangle \equiv \mid \Psi_0\rangle$. One can easily show that
this state in an eigenstate of the operator $\hat B_-$:
\begin{equation}
\hat B_-\mid z\rangle = z\mid z\rangle
\label{eqbcs}
\end{equation}
and satisfies the condition
\begin{equation}
(\hat B_- - z)\frac{\partial}{\partial z}\mid z\rangle = 
\mid z\rangle\,.
\label{eqbz}
\end{equation}
The state $\mid z\rangle$ coincides with the coherent state defined in
Ref.~\cite{ref8} using a generalized exponential function. When the
Lie algebra associated with the shape-invariant potential is SU(1,1)
\cite{ref3,ref7}, this is not the standard coherent state introduced
in \cite{ref9}, but the state introduced by Barut and Girardello
\cite{ref10}.

If a forced harmonic oscillator is in the ground state for $t=0$, it
evolves into the harmonic oscillator coherent state. We must emphasize
that the coherent states described here, in general, do not have such
a simple dynamical interpretation. To illustrate this point we
consider the time-dependent Hamiltonian
\begin{equation}
\hat h(t) = \hat B_+\hat B_- + f(t)\left[ {\rm
e}^{iR(a_1)t/\hbar}\hat B_+ +\hat B_- {\rm
e}^{-iR(a_1)t/\hbar}\right] \,,
\label{eqohf}
\end{equation}
where $f(t)$ is an arbitrary function of time. The solution of the
time-evolution equation 
\begin{equation}
i\hbar\frac{\partial\hat u(t)}{\partial t} = \hat h(t)\hat u(t)
\label{eqohe}
\end{equation}
can be written as 
\begin{equation}
\hat u(t) = \exp\left\{-\frac{i}{\hbar}\hat B_+ \hat B_- t\right\}\hat
u_I(t)\,.
\label{eqohe1}
\end{equation}
Substituting Eq.~(\ref{eqohe}) into Eq.~(\ref{eqohe1}) one can show
that $\hat u_I(t)$ satisfies the equation 
\begin{equation}
i\hbar\frac{\partial\hat u_I(t)}{\partial t} = f(t)\left[ \hat B_+ +
\hat B_-\right]\hat u_I(t)\,.
\label{eqdev}
\end{equation}
The solution of Eq.~(\ref{eqdev}) can be immediately written to be
\begin{equation}
\hat u_I(t) = \exp\left\{-\frac{i}{\hbar}\int_0^t
f(t^\prime)\,dt^\prime \left[\hat B_+ + \hat B_-\right]\right\} \,.
\label{eqdev1}
\end{equation}
Hence under the time-evolution the ground state evolves into the state
\begin{equation}
\mid \Psi,t\rangle = \hat u_I(t)\mid 0\rangle\,,
\label{eqpsit}
\end{equation}
which is not equivalent to the state given in Eq.~(\ref{eqcs}).

\section{Self-Similar Potentials and $q$-Algebras}

Shabat \cite{ref11} and Spiridonov \cite{ref12} discussed
reflectionless potentials with an infinite number of bound states.
These self-similar potentials are shown to be shape-invariant in
Ref.~\cite{ref6}. In this case the parameters are related by a
scaling:
\begin{equation}
a_n = q^{n-1}a_1\,.
\label{eqsc}
\end{equation}
Barclay {\it et al.} studied such shape-invariant potentials in detail
\cite{ref6}. In the simplest case studied by them the remainder of
Eq.~(\ref{eqsi}) is given by
\begin{equation}
R(a_1)= ca_1 \,,
\label{eqsc1}
\end{equation}
where $c$ is a constant and the operator introduced in
Eq.~(\ref{eqsio}) by
\begin{equation}
\hat T(a_1) = 
\exp{\left\{(\log q)a_1\frac{\partial}{\partial a_1}\right\}}\,.
\label{eqta}
\end{equation}
Hence the energy eigenvalue of the $n-$th excited state is 
\begin{eqnarray}
E_n &=& R(a_1) + R(a_2) + \dots + R(a_n) \nonumber \\
&=& (1 + q + q^2 + \dots + q^{n-1})ca_1 \nonumber \\ 
&=& \frac{1-q^n}{1-q}\,ca_1
\label{eqen}
\end{eqnarray}
which is the spectra of quantum oscillator \cite{ref13}. Introducing
the scaled operators
\begin{equation}
\hat K_\pm = \sqrt{q}\hat B_\pm
\label{eqjpm}
\end{equation}
one can show that the commutation relations of Eqs.~(\ref{eqcb1}),
(\ref{eqcb2}) and (\ref{eqcb3}) take the form
\begin{equation}
[\hat K_-,\hat K_+] = R(a_1)
\label{eqcj1}
\end{equation}
and
\begin{equation}
[\hat K_+,R(a_1)] =(q-1)R(a_1)\hat K_+\,.
\label{eqcKR}
\end{equation}
Note that the algebra associated with the self-similar potentials is
not a finite Lie algebra as $\hat K_+$ does not commute with
$R(a_1)\hat K_+^n$:
\begin{equation}
[\hat K_+,(q-1)^nR(a_1)\hat K_+^n] = 
(q-1)^{n+1}R(a_1)\hat K_+^{n+1} \,.
\label{eqcj2}
\end{equation}
Further introducing the operators 
\begin{equation}
\hat S_+ = \hat K_+R(a_1)^{-1/2}
\label{eqsma}
\end{equation}
and
\begin{equation}
\hat S_- = (\hat S_+)^\dagger = R(a_1)^{-1/2}\hat K_-\,,
\label{eqsme}
\end{equation}
using  Eq.~(\ref{eqcj1}) one can show that the standard $q$-deformed
oscillator relation is satisfied
\begin{equation}
\hat S_-\hat S_+ - q\hat S_+\hat S_- = 1\,.
\label{eqcsq}
\end{equation}

In the most general case for a self-similar potential the function
$W(x)$ of Eq.~(\ref{sup}) satisfies the condition \cite{ref11,ref12} 
\begin{equation}
  \label{eq:yy1}
  W(x)\,\buildrel{a_1\rightarrow a_2}\over{\longrightarrow}\, 
\sqrt{q}W (\sqrt{q}x)\,,
\end{equation}
or equivalently 
\begin{equation}
\label{eq:yy2}
\hat A^\dagger(x), \hat A (x) 
\,\buildrel{a_1 \rightarrow a_2}\over{\longrightarrow}\, 
\sqrt{q} \hat A^\dagger( \sqrt{q}x),  \sqrt{q} \hat A( \sqrt{q}x)\,.  
\end{equation}
Inserting Eq.~(\ref{eq:yy2}) into Eq.~(\ref{eqsi}) one obtains the
q-deformed form of  Eq.~(\ref{eqsi}) 
\begin{equation}
  \label{eq:yy3}
  \hat A(x)\hat A^\dagger(x) - q \hat A^\dagger( \sqrt{q}x) 
\hat A( \sqrt{q}x) = R(a_1) \,.
\end{equation}
Introducing the operators \cite{ell1}
\begin{equation}
  \label{eq:yy4}
  \hat C = \hat A(x) e^{-\frac{1}{2}p x \frac{d}{dx}}
\end{equation}
and 
\begin{equation}
  \label{eq:yy5}
  \hat C = e^{+\frac{1}{2}p x \frac{d}{dx}} \hat A^{\dagger}(x), 
\end{equation}
where $q=e^p$, Eq.~(\ref{eq:yy3}) can be rewritten as 
\begin{equation}
  \label{eq:yy6}
  \hat C\hat C^\dagger - q \hat C^\dagger \hat C = R(a_1) \,.
\end{equation}

Note that an algebraic approach to the self-similar potentials was
already introduced in Refs.~{\cite{ref7,ref7a}}. Here we would like to
establish that our algebra is identical to that in Ref.~\cite{ref7a}.
To this end we introduce
\begin{equation}
\hat J_3 =- \frac{1}{p} \log a_0 \,,
\label{y1}
\end{equation}
Using Eq.~(\ref{y1}), Eq.~(\ref{eqcb1}) can be written as 
\begin{equation}
  \label{eq:y2}
  [\hat B_-,\hat B_+] =  c \exp{(-p \hat J_3)}.
\end{equation}
Using Eq.~(\ref{eqsio}), one can show that for an arbitrary function
$f(a_n)$ of the parameters $a_n$ we can write 
\begin{equation}
  \label{eq:y3}
  f(a_n) \hat B_+ = \hat B_+ f(a_{n-1})
\end{equation}
and
\begin{equation}
  \label{eq:y4}
  f(a_n) \hat B_- = \hat B_- f(a_{n+1}).
\end{equation}
Using Eqs.~(\ref{eq:y3}) and (\ref{eq:y4}) one can easily prove the
commutation relation 
\begin{equation}
[\hat J_3, \hat B_\pm] = \pm \hat B_\pm\,.
\label{eqcj3}
\end{equation}
Eqs.~(\ref{eq:y2}) and (\ref{eqcj3}) represent the algebra introduced
in Ref.~{\cite{ref7a}}. This algebra is a deformation of the standard
$SO(2,1)$ algebra.  

The coherent state is easy to construct. The term multiplying $z^n$ in
Eq.~(\ref{eqcs}) is 
\begin{eqnarray}
z^n\hat B_-^{-n} \mid 0\rangle &=& z^n(\hat H^{-1}\hat B_+)^n
\mid 0\rangle \nonumber \\
&=& \left[ E_n(E_n-E_{n-1})(E_n-E_{n-2})\dots
(E_n-E_1)\right]^{-1/2}\mid n\rangle\,,
\label{eqznb}
\end{eqnarray}
where $\mid n\rangle$ is the short-hand notation for the $n$-th
excited state $\mid \Psi_n\rangle$ the energy of which is $E_n$.
Inserting Eq.~(\ref{eqen}) into Eq.~(\ref{eqznb}) one can write down
the coherent state as
\begin{equation}
\mid z\rangle = \sum_{n=0}^\infty \frac{z^n}{\sqrt{\left[R(a_1)
\right]^n}}\,\frac{(1-q)^{n/2}q^{-n(n-1)/4}}{\sqrt{(q;q)_n}}\mid 
n\rangle\,,
\label{eqcsq1}
\end{equation}
where the $q$-shifted factorial $(q;q)_n$ is defined as $(z;q)_0 = 1$
and $(z;q)_n = \prod_{j=0}^{n-1}(1-zq^j)\,,$ $n = 1,2,\dots$. One
observes that the norm of this state belongs to the one-parameter
family of $q$-exponential functions considered by Floreanini {\it et
  al.} \cite{ref14}. An alternative approach to the coherent states
for the q-algebras was given in Ref.~\cite{ref13} and was used to
construct path integrals in Ref.~\cite{ell2}.

\section*{ACKNOWLEDGMENTS}

This work was supported in part by the U.S. National Science
Foundation Grant No.\ PHY-9605140 at the University of Wisconsin, and
in part by the University of Wisconsin Research Committee with funds
granted by the Wisconsin Alumni Research Foundation.  M.A.C.R.\ 
acknowledges the support of Funda\c c\~ao de Amparo \`a Pesquisa do
Estado de S\~ao Paulo (Contract No.\ 96/3240-5). A.N.F.A. acknowledges
the support of Funda\c c\~ao Coordena\c c\~ao de Aperfei\c coamento de
Pessoal de N\'{\i}vel Superior (Contract No. BEX0610/96-8).

\end{document}